\documentclass[11pt,a4paper]{article}
\pdfoutput=1
\usepackage{jheppub}
\usepackage{amsmath,amssymb,amsthm,amsfonts}
\allowdisplaybreaks
\usepackage{slashed} 
\usepackage{graphicx,color,float}
\usepackage{subfigure}
\usepackage{dcolumn}
\usepackage{hyperref}      
\usepackage{bm}
\usepackage{epsfig}
\usepackage{epstopdf}
\usepackage{setspace}
\usepackage{comment}
\usepackage{multirow}
\usepackage{orcidlink}
\usepackage[normalem]{ulem}
\usepackage{fancyvrb}
\usepackage{xcolor} 

\title{Searching for long-lived particles from stopped pions and muons at the CiADS-BDE}

\author[a]{Zeren Simon Wang,\,\orcidlink{0000-0002-1483-6314}}
\emailAdd{wzs@mx.nthu.edu.tw}

\author[a,1]{Yu Zhang,\,\orcidlink{0000-0001-9415-8252}\note{Corresponding author}}
\emailAdd{dayu@hfut.edu.cn}

\author[b,c,d,2]{Liangwen Chen\,\orcidlink{0000-0002-1592-288X}\note{Corresponding author}}
\emailAdd{chenlw@impcas.ac.cn}

\affiliation[a]{School of Physics, Hefei University of Technology, Hefei 230601, China}
\affiliation[b]{Heavy Ion Science and Technology Key Laboratory, Institute of Modern Physics, Chinese Academy of Sciences, Lanzhou 730000, China}
\affiliation[c]{Advanced Energy Science and Technology Guangdong Laboratory, Huizhou 516000, China}
\affiliation[d]{School of Nuclear Science and Technology, University of Chinese Academy of Sciences, Beijing 101408, China}

\abstract{The CiADS-BDE is a beam-dump experiment recently proposed for searching for light, long-lived particles (LLPs) at China initiative Accelerator Driven System. Primarily thanks to the large numbers of protons on target at the experiment, it has been shown to be sensitive to large, unique regions of the parameter space of dark photon, with a small detector volume of $\mathcal{O}(0.01\text{--}1)$ m$^3$. Here, we explore the search prospect of the CiADS-BDE for a series of new-physic models predicting LLPs that could emanate from decays at rest of charged pions and muons at the facility, namely, heavy neutral leptons, axionlike particles, and light binos in the R-parity-violating supersymmetry. For these benchmark models, we find that the CiADS-BDE can also probe vast parameter regions beyond the existing bounds.}

\begin{document}
\maketitle

\section{Introduction}\label{sec:intro}

In recent years, searching for new physics beyond the Standard Model (BSM) in the form of so-called long-lived particles (LLPs) has become an increasingly important direction in high energy physics~\cite{Alimena:2019zri,Lee:2018pag,Curtin:2018mvb,Beacham:2019nyx}.
The LLPs, once produced, travel a macroscopic distance before decaying, possibly leading to exotic signatures at terrestrial experiments such as displaced vertices and missing energy.
LHC collaborations such as ATLAS and CMS have reported numerous LLP searches; see, for instance, Refs.~\cite{ATLAS:2024qoo,ATLAS:2024vnc,ATLAS:2024ocv,CMS:2024bvl,CMS:2024xzb,CMS:2024ita}.
Moreover, dedicated detectors have been proposed or even under operation at the LHC such as FASER~\cite{Feng:2017uoz,FASER:2018eoc,FASER:2022hcn} and MATHUSLA~\cite{Curtin:2018mvb,MATHUSLA:2020uve,Chou:2016lxi}.
In particular, FASER has published its initial results~\cite{FASER:2023tle,FASER:2024bbl} with data already collected during LHC Run3.
Besides the LHC, past and existing beam-dump experiments have obtained stringent bounds on LLPs predicted in various BSM physics scenarios; see e.g.~Refs.~\cite{CHARM:1983ayi,WA66:1985mfx,Bjorken:1988as}.

In a previous work~\cite{Chen:2024fzk}, we have proposed a beam-dump experiment to be operated at China initiative Accelerator Driven System (CiADS)~\cite{liu2017physics,Liu:2019cdo,Wang:2019ddc,Liu:2019xgd,He:2023izb,Cai:2023caf,Wang:2024joa}, called CiADS-BDE.
CiADS would be the first prototype Accelerator Driven System in the world to be operated at megawatt level, designed for developing technologies of nuclear waste disposal.
During beam commissioning at CiADS, a proton beam with energy of 600 MeV (2 GeV in the upgrade of CiADS) impinges on a beam dump (an oxygen-free copper target as currently planned), potentially producing LLPs.
The facility is currently under construction and planned to start operation in 2028.
Its number of protons on target (POTs) per year is expected to be in the order of magnitude $\mathcal{O}(10^{23})$, possibly allowing for relatively large production rates of LLPs.
We have studied the sensitivity reach of the CiADS-BDE, a cylindrical detector of length 1~m and radius 0.1~or 1~m, to be placed 10~m downstream the beam dump of the CiADS facility.
The space between the beam dump and the detector is available for implementation of veto and shielding, for the purpose of suppressing background events originating from the dump.
With liquid scintillator and Incom Large Area Picosecond Photodetectors (LAPPDs$^{\text{TM}}$) employed, the detector is able to reconstruct tracks and discriminate between Cherenkov and scintillation light emission.
In Ref.~\cite{Chen:2024fzk}, we have considered a benchmark model, the dark photon, which interacts with Standard-Model (SM) particles via its kinetic mixing with the SM U(1) gauge field.
The dominant production modes of the dark photon at the facility are decays of the $\pi^0$ and $\eta$ mesons and proton bremsstrahlung.
We have restricted ourselves to the signature of an electron-positron pair from the dark-photon decays.
Numerical results of Ref.~\cite{Chen:2024fzk} show promising search prospect, probing large parameter regions that are both currently allowed and hard to access by other future experiments.

In this work, we extend these efforts by focusing on a series of LLP candidates that can be produced from decays at rest of the $\pi^+$ mesons and the $\mu^+$ leptons at CiADS.
Dominantly owing to ionization and excitation effects of the atoms in the high-density target material, the $\pi^+$ mesons, after production, quickly lose energy as they propagate and come to rest.
Their usual decay products, the $\mu^+$ leptons, undergo the same energy-loss processes until coming to rest.\footnote{We note that, in comparison, negatively charged pions and muons are captured and absorbed by the atomic nuclei in the target material as a result of their Coulomb attraction to the positively charged nuclei.}
In a variety of BSM models, LLPs are predicted to be produced from decays of the muons and charged pions as long as kinematically allowed.
Thus, here, we study heavy neutral leptons (HNLs) that are SM singlets and can mix with active neutrinos of different flavors~\cite{Shrock:1980vy,Shrock:1980ct,Shrock:1981wq}, axionlike particles (ALPs) that can be either electrophilic~\cite{Altmannshofer:2022ckw} or charged-lepton-flavor-violating (LFV)~\cite{Heeck:2017xmg,Calibbi:2020jvd}, as well as light binos in the R-parity-violating supersymmetry~\cite{Barbier:2004ez,Dreiner:1997uz,Mohapatra:2015fua,deVries:2015mfw}.
Such models are well motivated for various reasons including explaining the non-zero active-neutrino masses, dark matter, hierarchy problem, among others.
As in Ref.~\cite{Chen:2024fzk}, we focus on the signature of an electron-positron pair arising from the displaced decays of the LLPs.
We perform Monte-Carlo (MC) simulations in this work in order to estimate the sensitivity reach of the proposed CiADS-BDE to these BSM scenarios, and we will show that the CiADS-BDE can probe unique parts in the parameter space of the considered models.

This paper is organized as follows.
In the next section, we introduce the experimental setup of the CiADS-BDE.
We then discuss the theoretical models we choose for numerical studies in Sec.~\ref{sec:model}, followed by Sec.~\ref{sec:results} where we present and elaborate on sensitivity results.
We conclude the work in Sec.~\ref{sec:conclusions} with a summary and an outlook.

\section{The beam-dump experimental setup and the simulation procedure}\label{sec:experiment}

In Ref.~\cite{Chen:2024fzk} we proposed a liquid-scintillator detector to be installed 10~m behind the beam dump at CiADS.
The detector is supposed to be able to reconstruct charged leptons including both energy and directionality, and thus reconstruct displaced vertices emerging inside the detector.
For the geometry of the cylindrical detector, we fix the length at 1~m and choose two possible radii of 0.1~m and 1~m, for numerical studies.
Its orientation along the barrel is aligned with the beam direction.
Following Ref.~\cite{Chen:2024fzk}, we consider two proton beam-energy setups, 600~MeV and 2~GeV, both with a current of 5~mA.
With an expected duty factor of 75\% (9 month per year for beam commissioning), the POTs per year are estimated to be $6.6\times 10^{23}$ with both beam-energy setups~\cite{Chen:2024fzk}.
Further, we will assume a 5-year operation duration.

The rate of $\pi^+$ decay at rest (DAR) per proton is a function of the proton beam energy, and is estimated with GEANT4 simulation~\cite{GEANT4:2002zbu} to be $2.4\%$ and $17.2\%$, for beam energy of 600 MeV and 2 GeV, respectively~\cite{Ciuffoli:2018qem}.
Furthermore, the anti-muons all stem from decays at rest of the $\pi^+$ mesons with a branching ratio of $\sim 99.99\%$~\cite{ParticleDataGroup:2024cfk} and we assume that they all decay at rest inside the beam dump.

As in Ref.~\cite{Chen:2024fzk}, we confine ourselves to the signal decay of the LLPs into an electron-positron pair (plus anything).
The implementation of liquid scintillator and LAPPDs$^{\text{TM}}$ allows for reconstruction of the electron-positron pair.
In order to reduce background level, we demand that both the electron and positron should have an energy of at least 17 MeV and their opening angle should be larger than 15$^\circ$, following Ref.~\cite{Soleti:2023hlr} where SHiNESS, a similar detector experiment, was proposed to be operated near the target at European Spallation Source (ESS)~\cite{Garoby:2017vew}.
Background events mimicking the di-lepton signal forming a displaced vertex stem from cosmic rays, neutrino charged-current interactions, neutrino neutral-current interactions, as well as the beam $\bar{\nu}_e$ component.
Following Ref.~\cite{Chen:2024fzk}, we take an order-of-magnitude estimate on the total background level, and assume there should be $100\pm 50$ background events per year.
Thus, for 5 years' operation, we conclude that an excess of 184.1 signal events would correspond to exclusion bounds at 90\% confidence level (C.L.).

We apply MC simulation techniques in order to determine the acceptance of the CiADS-BDE to the LLPs we choose to study.
We first simulate the decay of the mother particles ($\pi^+$ or $\mu^+$) into the LLP on hand plus other particles and then the decay of the LLPs into an electron-positron pair plus anything.
For the $i^{\text{th}}$-generated LLP, we check if the kinematic cuts on the charged leptons are satisfied; if so, we compute the probability of this LLP to decay inside the detector's fiducial volume ($\epsilon^{\text{decay in f.v.}}_{i^\text{th}\text{ LLP}}$) with exponential decay distribution functions taking into account the boosted decay length and the moving direction of the LLP.
This can be formulated in the following manner:
\begin{eqnarray}
	\epsilon^{\text{cut\&decay}}_{i^{\text{th}\text{ LLP}}} =
\begin{cases}
0, \text{ if failing the kinematic cuts,}\\
\epsilon^{\text{decay in f.v.}}_{i^\text{th}\text{ LLP}}, \text{ if passing the kinematic cuts,}
\end{cases}
\end{eqnarray}
where ``f.v.'' is for ``fiducial volume'', ``cut'' points to the kinematic cuts on the final-state leptons, and ``decay'' refers to the decay probability inside the fiducial volume.
In the end, we calculate the arithmetic average of the probability of the simulated LLPs to both pass the leptonic kinematic cuts and decay inside the CiADS-BDE, which corresponds to the acceptance $\epsilon^{\text{cut\&acc.}}_{\text{avg.}}$:
\begin{eqnarray}
	\epsilon^{\text{cut\&acc.}}_{\text{avg.}}=\frac{1}{N_{\text{sim.}}}\sum_{i=1}^{N_{\text{sim.}}} \epsilon^{\text{cut\&decay}}_{i^{\text{th}\text{ LLP}}} \,,
\end{eqnarray}
where $N_{\text{sim.}}$ denotes the total number of the simulated signal events.
We refer to Ref.~\cite{Chen:2024fzk} for more detail of this computation.
Finally, in order to estimate the signal-event numbers $N_S$, we include in the computation the number of the mother particles decaying at rest $N_{\pi^+/\mu^+}^{\text{DAR}}$, the decay branching ratio (BR) of the mother particle into the LLP, the acceptance $\epsilon^{\text{cut\&acc.}}_{\text{avg.}}$, as well as the decay BR of the LLP into the signal final states:
\begin{eqnarray}
    N_S = N_{\pi^+/\mu^+}^{\text{DAR}}\cdot \text{BR}(\pi^+/\mu^+\to \text{LLP}+X) \cdot \epsilon^{\text{cut\&acc.}}_{\text{avg.}} \cdot\text{BR}(\text{LLP}\to e^- e^+ +Y), 
\end{eqnarray}
where $X$ and $Y$ can be anything.

\section{Theoretical models}\label{sec:model}

In this section, we introduce the theoretical models we consider for sensitivity studies.

\subsection{Heavy neutral leptons}\label{subsec:model_hnl}

Observations of neutrino oscillation~\cite{Super-Kamiokande:1998kpq,deSalas:2017kay} have established the massive nature of the active neutrinos, constituting the first evidence of BSM physics.
The most popular explanation invokes the existence of so-called sterile neutrinos which are supposed to be SM singlets that mix with the active neutrinos via suppressed strengths.
Via seesaw mechanisms~\cite{Minkowski:1977sc,Yanagida:1979as,Gell-Mann:1979vob,Mohapatra:1979ia,Schechter:1980gr,Wyler:1982dd,Mohapatra:1986bd,Bernabeu:1987gr,Akhmedov:1995ip,Akhmedov:1995vm,Malinsky:2005bi}, the tiny but non-vanishing active-neutrino masses are explained.
In particular, in the context of collider physics, sterile neutrinos are often called heavy neutral leptons.
Here, for the purpose of performing a phenomenological study, we assume that the neutrinos are of Majorana nature and that there is only one HNL kinematically relevant which mixes with the active neutrinos of a single generation only.
We will consider two cases, with the HNL mixed with the electron neutrino and muon neutrino, respectively.
We write down the interaction Lagrangian between the HNL and the $W$- and $Z$-bosons, after electroweak symmetry breaking,
\begin{eqnarray}
\mathcal{L}_N \supset\frac{g}{\sqrt{2}}\sum_\alpha V_{\alpha N} \bar{l}_\alpha \gamma^\mu P_L N W^-_{L\mu}  + \frac{g}{2 \cos{\theta_W}}\sum_{\alpha,i}V^L_{\alpha i}V^*_{\alpha N} \bar{N} \gamma^\mu P_L \nu_i Z_\mu, \label{eqn:HNL_Lag}
\end{eqnarray}
where $g$ is the SU(2) gauge coupling, $V_{\alpha N}$ is the mixing-matrix element between the HNL and the active neutrino $\nu_\alpha$ with $\alpha=e,\mu,\tau$, $l_\alpha$ labels the SM charged leptons, $V^L_{\alpha i}$ is the PMNS mixing matrix with $i=1,2,3$, and $P_L=\frac{1-\gamma^5}{2}$ is the left-chiral projection operator.

For the case of the HNL mixed with the electron neutrino, i.e.~$|V_{eN}|^2\neq 0$, two signal processes are relevant: $\pi^+\to e^+ N, N\to \nu_e \, e^+ e^-$ and $\mu^+\to \bar{\nu}_\mu \, e^+ N, N\to \nu_e e^+ e^-$.
Our numerical results show that the latter process is sub-dominant, and therefore we will focus on the former for numerical studies.
For case of the HNL mixed with the muon neutrino only, we have instead $\mu^+\to \nu_e \, e^+ N, N\to \nu_\mu \, e^+ e^-$, mediated by $|V_{\mu N}|^2$.
For a detailed discussion of the decay rates involved in these processes, we refer to Refs.~\cite{Bondarenko:2018ptm,Ema:2023buz}.

\subsection{Electrophilic axionlike particles}\label{subsec:model_alp_e}

ALPs (and axions) are perhaps the most plausible solution to the strong CP problem~\cite{ParticleDataGroup:2024cfk,Peccei:1977ur,Peccei:2006as}, and additionally serve as dark-matter candidates~\cite{Dine:1982ah,Abbott:1982af,Preskill:1982cy,Marsh:2015xka,Lambiase:2018lhs,Auriol:2018ovo,Houston:2018vrf}.
Via the spontaneous breaking of an assumed approximate global U(1) Peccei-Quinn symmetry~\cite{Peccei:1977hh}, they emerge as the associated pseudo-Nambu-Goldstone bosons.
In contrast to the axions, the ALPs' mass and couplings to the SM particles are decoupled, allowing for more degrees of freedom in phenomenological studies.
Here, despite the possibility of the ALP coupled to various SM particles, we restrict ourselves to the scenario where the ALP is leptophilic, thus coupled to the charged leptons only.
Such a scenario is well motivated for relations to charged lepton flavor violation~\cite{Han:2020dwo,Cheung:2021mol,Bertuzzo:2022fcm}, leptonic electric dipole moment~\cite{Kirpichnikov:2020lws}, and the muon anomalous magnetic moment~\cite{Bauer:2019gfk,Cornella:2019uxs,Buen-Abad:2021fwq,Bauer:2021mvw,Davoudiasl:2024vje}.
We will assume an electrophilic ALP in this work, with weak-violating couplings~\cite{Altmannshofer:2022ckw}, for which a four-point vertex between the $W$-boson, electron, electron neutrino, and the ALP appears with no helicity-flip suppression.
We refer to Refs.~\cite{Altmannshofer:2022ckw,Lu:2022zbe,Buonocore:2023kna,Wang:2024zky,Jiang:2024cqj} for recent phenomenological studies on such electrophilic ALPs that can be either promptly decaying or long-lived.

The effective Lagrangian is given as
\begin{eqnarray}
    \mathcal{L}_{e\text{ALP}}\supset \partial_\mu a\, \frac{c_{ee}}{2\Lambda}\bar{e} \, \gamma^\mu \,\gamma_5\, e,\label{eqn:eALP_Lag1}
\end{eqnarray}
where $a$ is the ALP, $e$ denotes the electron, $c_{ee}$ is a dimensionless coupling, and $\Lambda$ is the effective cut-off scale.

After integration by part and equations of motion are employed, Eq.~\eqref{eqn:eALP_Lag1} is transformed to
\begin{eqnarray}
    \mathcal{L}_{e\text{ALP}}\supset i\frac{c_{ee} m_e}{\Lambda} a \, \bar{e} \, \gamma_5 \,e + \frac{ig}{2\sqrt{2}}\frac{c_{ee}}{\Lambda}a(\bar{e}\,  \gamma^\mu\, P_L\, \nu_e)W_\mu^-+\cdots+\text{h.c.}, \label{eqn:eALP_Lag2}
\end{eqnarray}
where $m_e$ is the electron mass, and the dots include terms from chiral anomaly which are irrelevant to our phenomenology.

The two terms in Eq.~\eqref{eqn:eALP_Lag2} both mediate the production of the ALP in $\pi^+\to e^+ \nu_e \, a$ decays, where, in particular, the second term is unsuppressed by helicity flip and is therefore dominant.
The decay-rate formulas for $\pi^+\to e^+ \nu_e \, a$ can be found in Ref.~\cite{Altmannshofer:2022ckw}.
Further, the first term of Eq.~\eqref{eqn:eALP_Lag2} induces our signal decay of the ALP into an electron-positron pair, for which the decay width $\Gamma(a\to e^+ e^-)$ is computed with the following formula~\cite{Bauer:2017ris,Bauer:2018uxu,Bauer:2021mvw,Lu:2022zbe},
\begin{eqnarray}
    \Gamma(a\to e^+ e^-) = \frac{c_{ee}^2}{8\pi \Lambda^2} m_e^2 m_a \sqrt{1 - \frac{4 m^2_{e}}{m_a^2}},
\end{eqnarray}
with $m_a$ denoting the ALP mass.
The ALP in the mass range of our interest decays into an electron-positron pair essentially with a 100\% branching ratio, as the loop-induced decay channel into a photon pair is strongly suppressed~\cite{Cheung:2021mol,Lu:2022zbe,Jiang:2024cqj}.

\subsection{Lepton-flavor-violating axionlike particles}\label{subsec:model_alp_lfv}

We study one further ALP scenario where the ALP is both leptophilic and charged-LFV.
In the SM of particle physics, flavor-changing-neutral-current interactions are absent at tree-level and are thus strongly suppressed.
This implies that observations of LFV processes can easily be an indication of BSM physics.
For instance, in the muon sector, the branching ratios of both $\mu\to e \gamma$ and $\mu\to 3e$ processes are predicted in the SM to be in the order of magnitude $\mathcal{O}(10^{-55})$~\cite{Petcov:1976ff,Hernandez-Tome:2018fbq} while experimental bounds are currently only at $\mathcal{O}(10^{-13})$~\cite{MEG:2016leq} and $\mathcal{O}(10^{-12})$~\cite{SINDRUM:1987nra}, respectively.
Even future experiments MEG II and Mu3e can improve these bounds by only up to four orders of magnitude~\cite{Baldini:2013ke,Blondel:2013ia}.
Moreover, ALPs with LFV couplings can explain anomalies in both electron and muon anomalous magnetic moments simultaneously~\cite{Bauer:2019gfk}.

We work with the following effective Lagrangian,
\begin{eqnarray}
    \mathcal{L}^{\text{LFV}}_{\text{ALP}} \supset i\,g_{\mu e}\, a \,\bar{\mu}\,\gamma_5\, e + i\, g_{e e} \, a \, \bar{e} \,\gamma_5\, e,  \label{eqn:LFVALP_Lag}
\end{eqnarray}
where $g_{\mu e}$ and $g_{ee}$ are dimensionless couplings.
These interactions can induce our signal decays $\mu^+\to e^+ a$ and $a\to e^+ e^-$ with the couplings $g_{\mu e}$ and $g_{ee}$, respectively.
The decay widths of these processes can be calculated with the following formulas~\cite{Bauer:2017ris,Bauer:2018uxu,Bauer:2021mvw,Lu:2022zbe},
\begin{eqnarray}
    \Gamma(\mu^+\to e^+ \, a)&=& \frac{g_{\mu e}^2 m_\mu}{16\pi}\Big( 1 - \frac{m_a^2}{m_\mu^2}  \Big)^2, \\
    \Gamma(a \to e^+ e^-)    &=&  \frac{g_{e e}^2 m_a}{8\pi}   \sqrt{1-\frac{4m^2_e}{m_a^2}},  
\end{eqnarray}
where $m_{\mu/e/a}$ denotes the mass of the muon/electron/ALP.
The ALPs, considered to be produced from muon decays, decay exclusively into an electron-positron pair.

We note that although we work within an effective framework, ultra-violet (UV)-complete models predicting ALPs with LFV interactions include Dine-Fischler-Srednicki-Zhitnitsky (DFSZ) axion~\cite{Zhitnitsky:1980tq,Dine:1981rt}, leptonic familon~\cite{Froggatt:1978nt}, and majoron~\cite{Ibarra:2011xn}.
Further existing studies on LFV ALP can be found in e.g.~Refs.~\cite{Heeck:2017xmg,Calibbi:2020jvd,Cheung:2021mol,Panci:2022wlc,Zhang:2023vva,Calibbi:2024rcm,Knapen:2024fvh}.

\subsection{Light binos in the R-parity-violating supersymmetry}\label{subsec:model_bino}

In supersymmetric (SUSY) models, a $Z_2$ R-parity is often imposed, in order to ensure that the lightest supersymmetric particle (LSP) is stable.
However, a priori this is not necessarily the case and the R-parity can be broken.
Compared to the usual R-parity-conserving SUSY, the R-parity-violating SUSY (RPV-SUSY)~\cite{Barbier:2004ez,Dreiner:1997uz,Mohapatra:2015fua} can in addition explain the non-zero active-neutrino masses~\cite{Hall:1983id,Grossman:1998py,Hirsch:2000ef,Dreiner:2006xw,Dreiner:2011ft} and provide a more involving phenomenology at colliders~\cite{Dreiner:1991pe,deCampos:2007bn,Dercks:2017lfq,Dreiner:2023bvs}.
Furthermore, the RPV-SUSY can accommodate $B$-anomalies~\cite{Trifinopoulos:2019lyo,Hu:2020yvs,BhupalDev:2021ipu}, the ANITA anomaly~\cite{Collins:2018jpg,Altmannshofer:2020axr}, as well as the muon $g-2$ anomaly~\cite{Hu:2019ahp,Zheng:2021wnu,BhupalDev:2021ipu}.
Here, we consider the lightest neutralino $\tilde{\chi}^0_1$ as a bino~\cite{Gogoladze:2002xp,Dreiner:2009ic} in the RPV-SUSY that can be as light as below the GeV-scale or even massless~\cite{Domingo:2022emr}, and assume the light bino to be the LSP.
In this case, if we lift the GUT-inspired relation between the gauginos $M_1\approx 0.5\,M_2$~\cite{Choudhury:1995pj,Choudhury:1999tn}, the light bino decays (which is naturally the case in the RPV-SUSY) to avoid overclosing the Universe~\cite{Bechtle:2015nua}, and the light bino is not the dark matter~\cite{Belanger:2002nr,Hooper:2002nq,Bottino:2002ry,Belanger:2003wb,AlbornozVasquez:2010nkq,Calibbi:2013poa}, such a light bino is allowed by all laboratory~\cite{Dreiner:1991pe, Choudhury:1999tn, Dreiner:2009er,Dreiner:2009ic,Gogoladze:2002xp,Dreiner:2022swd}, cosmological and astrophysical bounds~\cite{Grifols:1988fw,Ellis:1988aa,Lau:1993vf,Dreiner:2003wh,Dreiner:2013tja,Profumo:2008yg,Dreiner:2011fp}.
Since the RPV couplings are bounded from above, the light binos are expected to be long-lived.
Existing phenomenological studies on searches for long-lived light binos in the RPV-SUSY at various experimental facilities can be found in, for instance, Refs.~\cite{Choudhury:1999tn,deVries:2015mfw,Dercks:2018eua,Wang:2019orr,Gehrlein:2021hsk,Dreiner:2023gir,Wang:2024zky}.
These works show that the light bino can be produced in decays of various mesons or the SM $Z$-boson (via small components of Higgsinos in the lightest neutralino), and here, we will focus on three benchmark scenarios of the RPV couplings with lepton number violation (LNV) where the light binos originate in decays of the charged pions or muons.

We write down the RPV superpotential here, where only trilinear LNV operators are considered:
\begin{eqnarray}
    W_{\text{RPV}}\supset \frac12 \lambda_{ijk} L_i L_j \bar{E}_k + \lambda'_{ijk} L_i Q_j \bar{D}_k. 
 \label{eqn:RPVbino_Lag}
\end{eqnarray}
Here, $i,j,k,=1,2,3$ are generation indices, and $\lambda_{ijk}$ and $\lambda'_{ijk}$ are dimensionless couplings.
The factor $\frac{1}{2}$ on the $\lambda_{ijk} L_i L_j \bar{E}_k$ operators are due to the anti-symmetry property of $\lambda_{ijk}$ such that $\lambda_{ijk}=-\lambda_{jik}$.

We list here the benchmark scenarios of the LNV RPV couplings we choose to study, together with the signal decay processes,
\begin{enumerate}
    \item $\mathbf{B1}$: $\lambda'_{111}$ and $\lambda_{131}$, with $\pi^+\overset{\lambda'_{111}}{\to} e^+ \tilde{\chi}^0_1, \tilde{\chi}^0_1\overset{\lambda_{131}}{\to} \overset{(-)}{\nu}_\tau \, e^- e^+$,
    \item $\mathbf{B2}$: $\lambda_{312}$ and $\lambda_{131}$, with $\mu^+\overset{\lambda_{312}}{\to} e^+ \tilde{\chi}^0_1 \, \nu_\tau, \tilde{\chi}^0_1   \overset{\lambda_{131}}{\to} \overset{(-)}{\nu}_\tau \, e^- e^+$,
    \item $\mathbf{B3}$: $\lambda_{121}$, with $\mu^+ \overset{\lambda_{121}}{\to} e^+ \tilde{\chi}^0_1 \, \bar{\nu}_e, \tilde{\chi}^0_1 \overset{\lambda_{121}}{\to} \overset{(-)}{\nu}_\mu  \, e^- e^+$.
\end{enumerate}
We note that in $\mathbf{B1}$ and $\mathbf{B2}$ there are two non-vanishing couplings that mediate the production and decay of $\tilde{\chi}^0_1$, respectively, while in $\mathbf{B3}$ a single coupling $\lambda_{121}$ induces both the production and decay.
The decay rates of these processes are suppressed by the heavy sfermion masses and proportional to $\Big(\frac{\lambda^{(')}}{m^2_{\tilde{f}}}  \Big)^2$ where $m_{\tilde{f}}$ denotes the mass of the associated sfermions.
In this work, we assume degenerate sfermion masses for simplicity.
To calculate $\Gamma(\pi^+ \overset{\lambda'_{111}}{\to} e^+ \tilde{\chi}^0_1)$ we make use of the analytic formulas provided in Ref.~\cite{deVries:2015mfw}, while the decay rates of the $\mu^+$ lepton and the light bino via the $\lambda$ couplings are computed with the procedures spelled out in Ref.~\cite{Dreiner:2023gir}.

\section{Numerical results}\label{sec:results}

We present numerical results in this section, in the form of contour curves of 184.1 signal events which correspond to exclusion limits at 90\% C.L. for 5-year data collection~(see the discussion in Sec.~\ref{sec:experiment}).

\subsection{Heavy neutral leptons}\label{subsec:results_hnl}

\begin{figure}[t]
	\centering
	\includegraphics[width=0.495\textwidth]{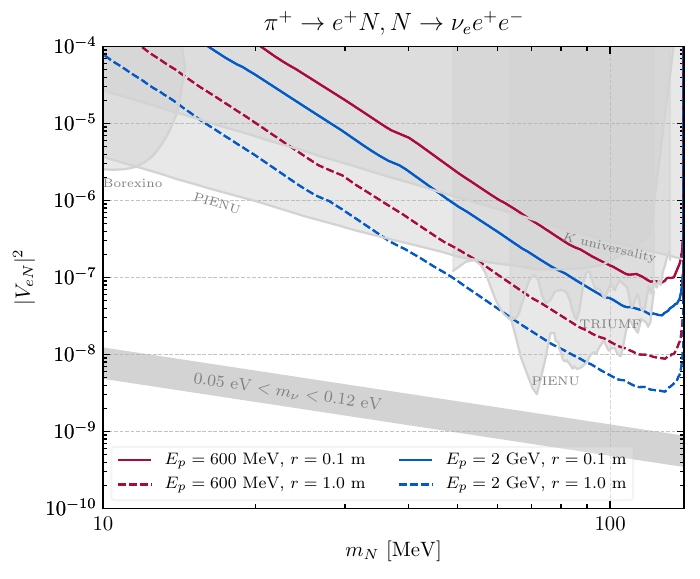}
	\includegraphics[width=0.495\textwidth]{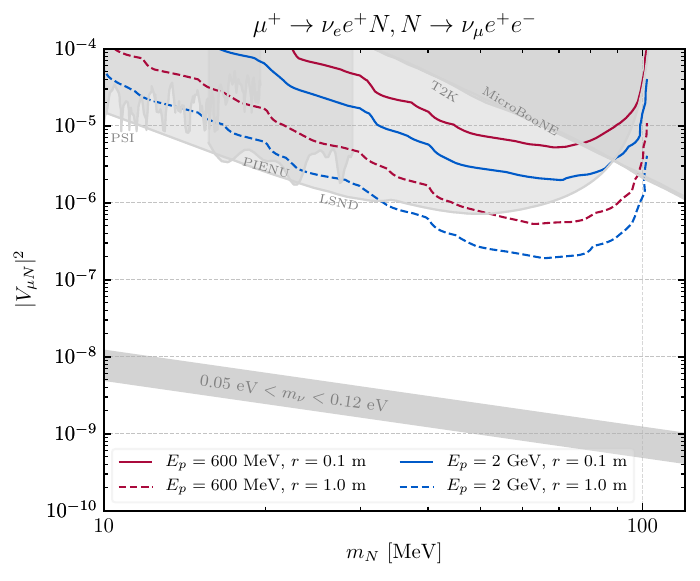}
	\caption{\textit{Left panel:} sensitivity reach of the CiADS-BDE to $|V_{eN}|^2$ as functions of $m_N$, considering HNLs produced in $\pi^+$ decays and 5-year data collection. The red (blue) lines are for proton-beam kinetic energy of 600 MeV (2 GeV), and the solid (dashed) lines are for the detector radius of 0.1 m (1.0 m). The light-gray area has been excluded by the Borexino~\cite{Borexino:2013bot}, PIENU~\cite{PIENU:2017wbj,Bryman:2019bjg}, and TRIUMF~\cite{Britton:1992xv} experiments, as well as lepton-flavor-universality considerations in kaon decays~\cite{NA62:2012lny,Bryman:2019ssi}.
    The dark-gray band covers the parameter region that could accommodate the non-zero active-neutrino masses via the canonical type-I seesaw relation $|V_{eN}|^2\sim m_{\nu}/m_N$ for active-neutrino mass between 0.05 eV~\cite{Canetti:2010aw} and 0.12 eV~\cite{Planck:2018vyg}.
    \textit{Right panel:} sensitivity reach of the CiADS-BDE to the sterile-neutrino mixing with muon neutrinos, as functions of $m_N$.  The upper gray region corresponds to the present bounds obtained at the PSI~\cite{Daum:1987bg}, PIENU~\cite{PIENU:2019usb}, LSND~\cite{LSND:2001akn,Ema:2023buz}, T2K~\cite{T2K:2019jwa,Arguelles:2021dqn}, and MicroBooNE~\cite{MicroBooNE:2021sov,Kelly:2021xbv} experiments.
}
 \label{fig:results_hnl}
\end{figure}

We show the sensitivity reach of the CiADS-BDE to the HNLs in Fig.~\ref{fig:results_hnl}, presented in the planes $|V_{eN}|^2$ vs.~$m_N$ (left plot) and $|V_{\mu N}|^2$ vs.~$m_N$ (right plot).
The results for the case of $E_p=600$ MeV ($E_p=2$ GeV) are plotted in red (blue), and those for the detector radius of $r=0.1$ m ($r=1.0$ m) are in solid (dashed) line style.
The same style will be taken for the sensitivity results of the other theoretical models.

In the left plot, the results are obtained by considering HNLs produced in the $\pi^+$ decay only; the $\mu^+$ decay mode is not shown since its sensitivity is almost always sub-dominant.
The existing bounds are shown as the gray-shaded region in the background, obtained at the Borexino~\cite{Borexino:2013bot}, PIENU~\cite{PIENU:2017wbj,Bryman:2019bjg}, and TRIUMF~\cite{Britton:1992xv} experiments, as well as by considering lepton-flavor-universality in kaon decays~\cite{NA62:2012lny,Bryman:2019ssi}.
The dark-gray band covers the parameter region that could explain the active-neutrino masses between 0.05 eV~\cite{Canetti:2010aw} and 0.12 eV~\cite{Planck:2018vyg} by the vanilla type-I seesaw relation $|V_{eN}|^2\sim m_\nu/m_N$.
We find that the detector setup with the smaller radius can probe only a small corner of the parameter space beyond the current bounds, while the setup with $r=1.0$ m can exclude values of $|V_{eN}|^2$ down to the order of magnitude $\mathcal{O}(10^{-9})$ for HNL mass just below the pion threshold, which is up to about 2 orders of magnitude beyond the current limits.
The $E_p=2$ GeV case is expected to perform better than the $E_p=600$ MeV case, because of the larger proportion of the charged pions' decay at rest at the CiADS beam dump.

In the right panel, the existing limits stem from the PSI~\cite{Daum:1987bg}, PIENU~\cite{PIENU:2019usb}, LSND~\cite{LSND:2001akn,Ema:2023buz}, T2K~\cite{T2K:2019jwa,Arguelles:2021dqn}, and MicroBooNE~\cite{MicroBooNE:2021sov,Kelly:2021xbv} experiments.
Here, we observe that the smaller version of the CiADS-BDE is insensitive to any parameter region beyond the present bounds, while the the larger version with $r=1.0$ m can probe new parameter regions for $m_N$ roughly between 0.04 GeV and 0.1 GeV, down to the order of magnitude $\mathcal{O}(10^{-7})$ for $|V_{\mu N}|^2$.

If compared to the proposed SHiNESS experiment~\cite{Soleti:2023hlr}, we find that the CiADS-BDE is sensitive to similar regions in the parameter space.

\subsection{Electrophilic axionlike particles}\label{subsec:results_alp_e}

\begin{figure}[t]
	\centering
	\includegraphics[width=0.495\textwidth]{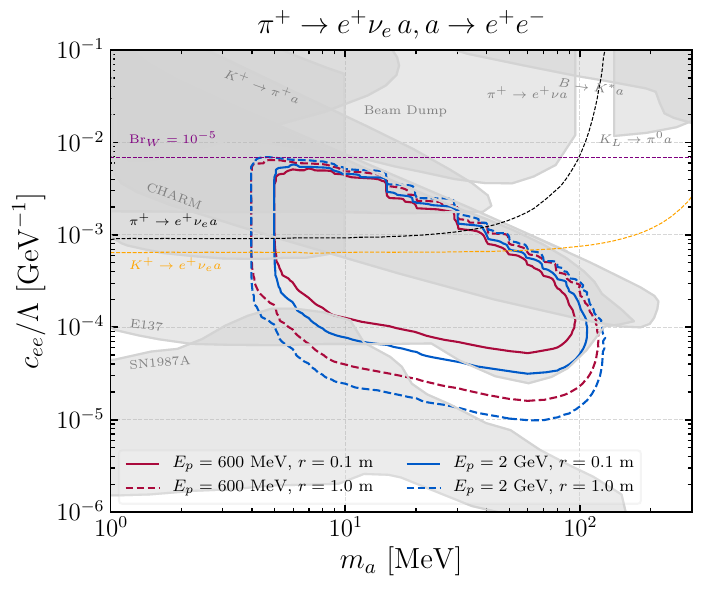}
	\caption{Sensitivity reach of the CiADS-BDE to the electrophilic ALP shown in the plane $c_{ee}/\Lambda$ vs.~$m_a$. The present constraints are shown in the light-gray regions, attained from electron beam-dump experiments~\cite{Konaka:1986cb,Riordan:1987aw,Bjorken:1988as,Bross:1989mp}, observation of supernova SN1987A~\cite{Carenza:2021pcm}, and searches for various flavor-changing-neutral-current decays of kaons and $B$-mesons~\cite{KTeV:2003sls,LHCb:2015ycz,NA62:2020xlg}. Bounds from searches for leptonic decays of $\pi^+$~\cite{SINDRUM:1986klz} and $K^+$~\cite{Poblaguev:2002ug} are also shown in gray. In addition, bounds from the proton beam-dump experiment CHARM~\cite{CHARM:1985anb} are displayed.
    Projections of future bounds at the PIONEER experiment (black dashed)~\cite{PIONEER:2022yag}, kaon factories (orange dashed)~\cite{Goudzovski:2022vbt}, and the LHC (purple dashed)~\cite{Altmannshofer:2022ckw}, are also extracted from Ref.~\cite{Altmannshofer:2020axr} and displayed here.
}
 \label{fig:results_alp_e}
\end{figure}

In Fig.~\ref{fig:results_alp_e}, we display the sensitivity reach of the CiADS-BDE to the electrophilic ALP, shown in the $(m_a, c_{ee}/\Lambda)$ plane.
The present bounds are shown in light gray, which are obtained from various laboratory searches~\cite{Konaka:1986cb,Riordan:1987aw,Bjorken:1988as,Bross:1989mp,KTeV:2003sls,LHCb:2015ycz,NA62:2020xlg,SINDRUM:1986klz,Poblaguev:2002ug,CHARM:1985anb} and astrophysical observations~\cite{Carenza:2021pcm}.
We also reproduce from Ref.~\cite{Altmannshofer:2022ckw} expected future bounds at the PIONEER experiment (black)~\cite{PIONEER:2022yag}, kaon factories (orange)~\cite{Goudzovski:2022vbt}, as well
as the LHC (purple)~\cite{Altmannshofer:2022ckw}, shown in dashed curves.

The results show that the CiADS-BDE with $r=1.0$ m can exclude an unique part of the parameter region that is currently unexcluded and hard to access at other future experiments.
Just as in the HNL case, here in the electrophilic-ALP case the CiADS-BDE is found to have sensitivity to a similar parameter region to the SHiNESS experiment~\cite{Wang:2024zky}.

\subsection{Lepton-flavor-violating axionlike particles}\label{subsec:results_alp_lfv}

\begin{figure}[t]
	\centering
	\includegraphics[width=0.495\textwidth]{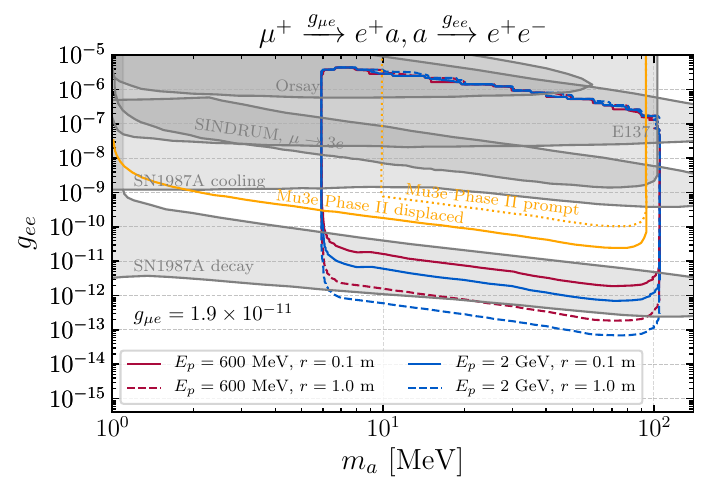}
	\includegraphics[width=0.495\textwidth]{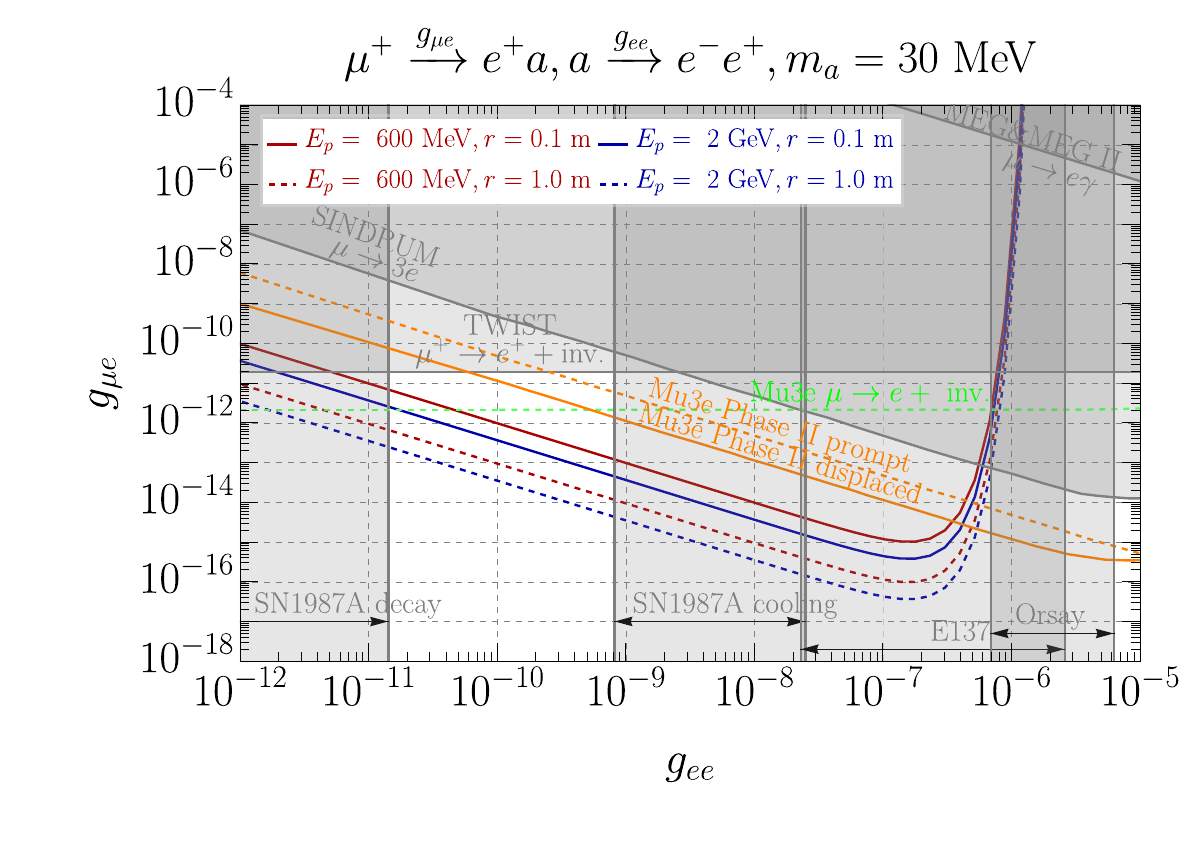}
	\caption{\textit{Left panel:} sensitivity reach of the CiADS-BDE to the LFV ALP, displayed in the $(m_a, g_{ee})$ plane, where we have fixed $g_{\mu e}=1.9\times 10^{-11}$ saturating the bounds from TWIST~\cite{TWIST:2014ymv}. The other relevant exiting bounds are shown in the gray regions, obtained from electron beam-dump experiments E137~\cite{Bjorken:1988as} and Orsay~\cite{Davier:1989wz} (and their reinterpretations~\cite{Liu:2016qwd,Liu:2017htz}), SINDRUM~\cite{SINDRUM:1987nra,SINDRUM:1986klz}, and SN1987A~\cite{Ferreira:2022xlw}.
    The expected sensitivities at the Phase II of the Mu3e experiment is overlapped here in the orange solid (dotted) line for a displaced-vertex (prompt) search strategy for the identical signal process, reproduced from Ref.~\cite{Knapen:2024fvh}.
    \textit{Right panel:} sensitivity reach shown in the plane $g_{\mu e}$ vs.~$g_{ee}$, for $m_a=30$ MeV.
    Besides the current bounds displayed in the left panel, here we present the combined limit from MEG and MEG II~\cite{MEG:2016leq,MEGII:2023ltw} on BR$(\mu\to e \gamma)$.
    Moreover, the Mu3e sensitivities are again extracted from Ref.~\cite{Knapen:2024fvh}, for both a $\mu\to e+\text{ inv.}$ search at the Phase I (green dashed line) and a prompt/displaced-vertex search for the same signal process at Phase II (orange dashed/solid line).
}
 \label{fig:results_alp_lfv}
\end{figure}

In Fig.~\ref{fig:results_alp_lfv} we present the sensitivity results for the LFV ALPs, in the plane $g_{ee}$ vs.~$m_a$ (left panel) and $g_{\mu e}$ vs.~$g_{ee}$ (right panel).
In the left plot, we fix $g_{\mu e}$ at $1.9\times 10^{-11}$ saturating the existing bounds on $\mu^+\to e^+ +$ invisible from the TWIST collaboration~\cite{TWIST:2014ymv}, thus maximizing the production rates of the ALP from $\mu^+$ decays.
The existing limits originate from laboratory experiments E137~\cite{Bjorken:1988as}, Orsay~\cite{Davier:1989wz}, and their reinterpretations~\cite{Liu:2016qwd,Liu:2017htz}, in addition to SINDRUM~\cite{SINDRUM:1987nra,SINDRUM:1986klz} and astrophysical observation of SN1987A~\cite{Ferreira:2022xlw}.
Furthermore, we display the expected sensitivities at the Phase II of the Mu3e experiment~\cite{Mu3e:2020gyw} with $5\times 10^{16}$ muon decays extracted from Ref.~\cite{Knapen:2024fvh}, where the orange solid and dotted lines are the results with a displaced-vertex and prompt search strategy, respectively, for the same signal process as what we study.
These findings show that with all the energy and detector-radius setups we have studied, the CiADS-BDE can probe a large part of the parameter space that is exactly between the regions that SN1987A cooling and SN1987A decay are sensitive to.
Compared to the Mu3e experiment, the CiADS-BDE clearly has an advantage, probing orders-of-magnitude smaller values of $g_{ee}$.

In the right panel, we fix the ALP mass at 30 MeV and show sensitivity reach of the CiADS-BDE in the $(g_{ee}, g_{\mu e})$ plane.
Existing bounds are shown in gray, obtained from the same experiments and observations mentioned above, in addition to the bounds from MEG and MEG II~\cite{MEG:2016leq,MEGII:2023ltw} measurement on BR$(\mu\to e \gamma)$.
The sensitivities of Mu3e is again reproduced from Ref.~\cite{Knapen:2024fvh}, where the green dashed line corresponds to a search for $\mu\to e+\text{ inv.}$ with Phase I of Mu3e ($2.5\times 10^{15}$ muon decays), and the orange dashed (solid) line is for a Phase-II prompt (displaced-vertex) search.
We find that for $g_{ee}$ roughly between $10^{-11}$ and $10^{-9}$, the CiADS-BDE can exclude $g_{\mu e}$ values down to the order of magnitude $\mathcal{O}(10^{-15})$ which is almost four orders of magnitude beyond the current bounds.
In contrast to Mu3e, the CiADS-BDE is sensitive to relatively smaller values of $g_{ee}$ and $g_{\mu e}$.

\subsection{Light binos in the R-parity-violating supersymmetry}\label{subsec:results_bino}

\begin{figure}[t]
	\centering
	\includegraphics[width=0.495\textwidth]{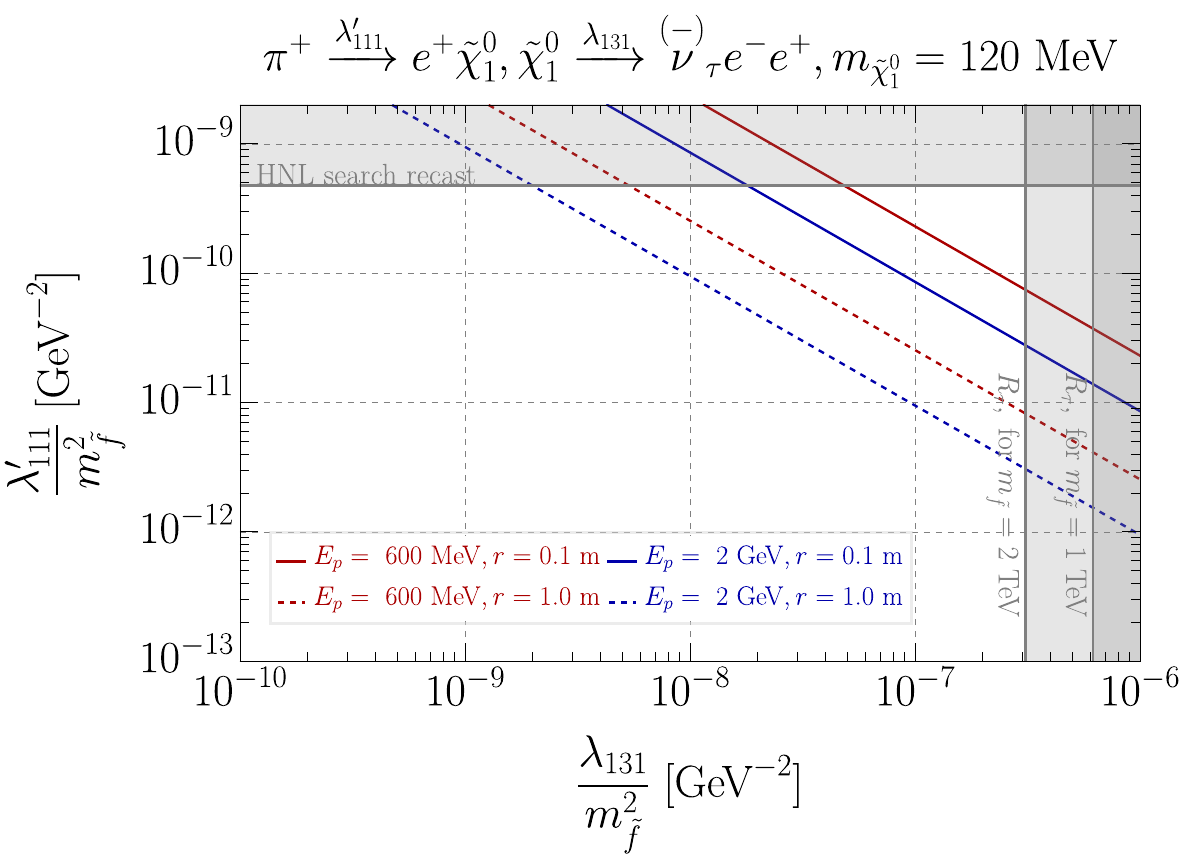}
	\includegraphics[width=0.495\textwidth]{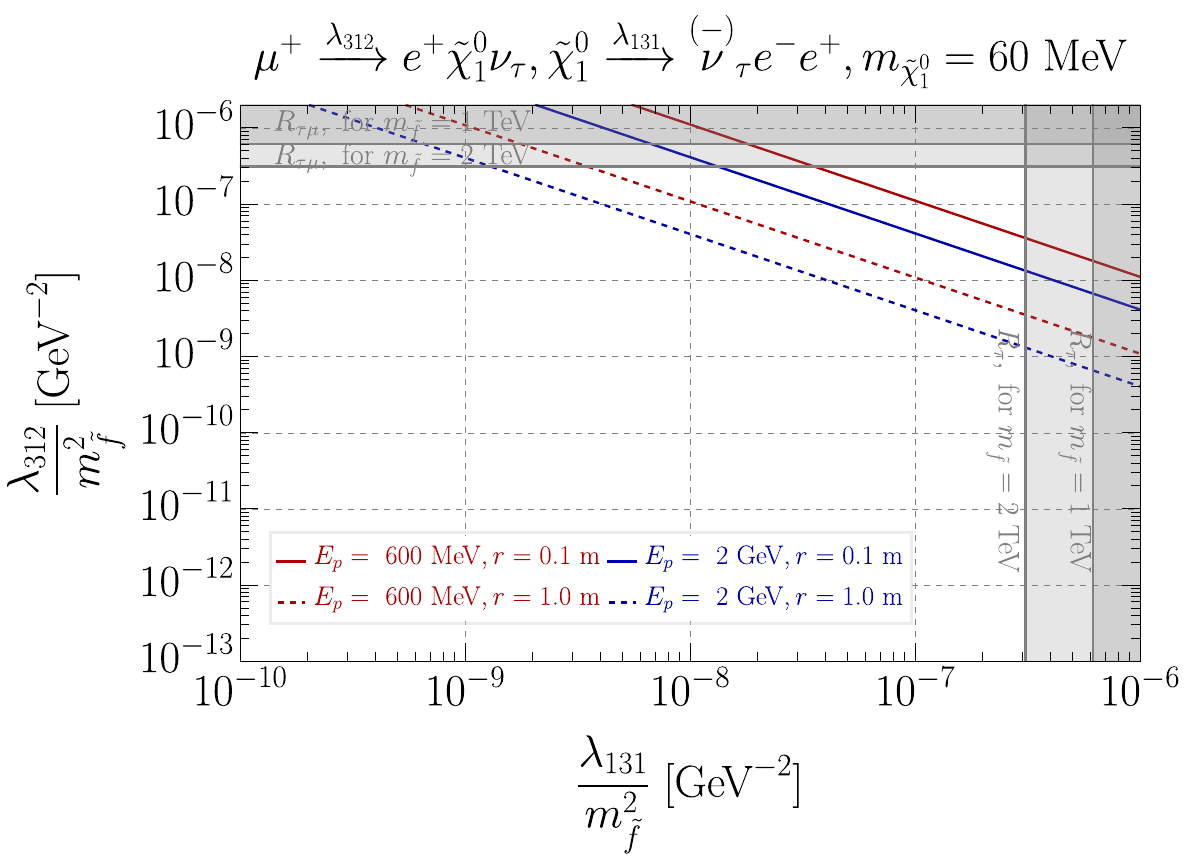}
	\includegraphics[width=0.495\textwidth]{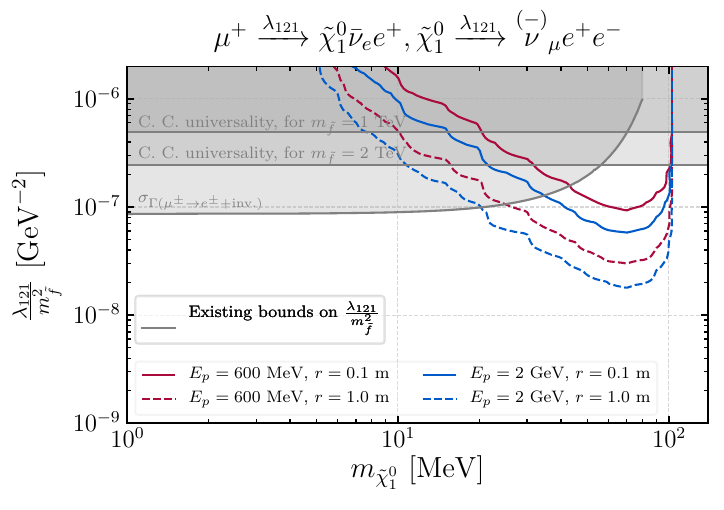}
	\caption{\textit{Upper-left panel:} sensitivity reach of the CiADS-BDE to the light bino, displayed in the plane $\lambda'_{111}/m^2_{\tilde{f}}$ vs.~$\lambda_{131}/m^2_{\tilde{f}}$, for $m_{\tilde{\chi}^0_1}=120$ MeV. The strongest bound on $\lambda'_{111}/m^2_{\tilde{f}}$ for the considered bino mass stems from recast of HNL searches~\cite{Dreiner:2023gir}. The present limits on $\lambda_{131}/m^2_{\tilde{f}}$ were derived in Ref.~\cite{Barger:1989rk} by considering $R_\tau$ (defined in the text), for which we assume two benchmark sfermion masses 1 TeV and 2 TeV.
                  \textit{Upper-right panel:} sensitivity reach in the plane $\lambda_{312}/m^2_{\tilde{f}}$ vs.~$\lambda_{131}/m^2_{\tilde{f}}$ for $m_{\tilde{\chi}^0_1}=60$ MeV. The upper bounds on $\lambda_{312}/m^2_{\tilde{f}}$ were attained in Ref.~\cite{Ledroit:1998} by considering measurements on $R_{\tau\mu}$ (see the text), where we take two benchmark sfermion masses 1 TeV and 2 TeV.
                  \textit{Lower panel:} sensitivity reach in the $(m_{\tilde{\chi}^0_1}, \lambda_{121}/m^2_{\tilde{f}})$ plane, where the present bounds were derived from measurements on C.~C.~universality~\cite{Barger:1989rk} and from uncertainty on the decay width of $\mu\to e+\text{invisible}$~\cite{Dreiner:2023gir}.
}
\label{fig:results_bino}
\end{figure}

The sensitivity results of the CiADS-BDE to the light binos in the RPV-SUSY are given in Fig.~\ref{fig:results_bino}, where we present three plots corresponding to the three benchmarks we choose to study, $\textbf{B1}, \textbf{B2},$ and $\textbf{B3}$, respectively.

For $\textbf{B1}$ (upper-left plot), we have fixed the bino mass $m_{\tilde{\chi}^0_1}$ at 120 MeV, and show the results in the $(\lambda_{131}/m^2_{\tilde{f}}, \lambda'_{111}/m^2_{\tilde{f}})$ plane.
For the fixed bino mass, the strongest bound on $\lambda'_{111}/m^2_{\tilde{f}}$ stems from recast of HNL searches~\cite{Dreiner:2023gir}, which is about $4.8\times 10^{-10}$ GeV$^{-2}$.
The leading bounds on $\lambda_{131}$, on the other hand, were derived from measurements on $R_\tau=\Gamma(\tau\to e\nu \bar{\nu})/\Gamma(\tau\to\mu \nu\bar{\nu})$~\cite{Barger:1989rk}: $\lambda_{131}<0.062\cdot \frac{m_{\tilde{e}_R}}{100\text{ GeV}}$.
Given the scaling of the bound on $\lambda_{131}$ with respect to the sfermion mass $m_{\tilde{e}_R}$, we show two vertical lines for sfermion masses of 1 TeV and 2 TeV, respectively.
Here, we find that the CiADS-BDE with any beam-energy and detector-radius combinations we choose can probe new parameter space beyond the present bounds.
In particular, for $E_p=2$ GeV and $r=1.0$ m, the proposed facility can exclude values of the considered RPV couplings by up to two orders of magnitude.

The sensitivity results of $\textbf{B2}$ are shown in the upper-right plot, in the plane $\lambda_{312}/m^2_{\tilde{f}}$ vs.~$\lambda_{131}/m^2_{\tilde{f}}$ with $m_{\tilde{\chi}^0_1}=60$ MeV.
Here, $\lambda_{312}$ and $\lambda_{131}$ are assumed to be non-vanishing, mediating the decays $\mu^+\to e^+ \tilde{\chi}^0_1 \nu_\tau$ and $\tilde{\chi}^0_1\to \overset{(-)}{\nu}_\tau e^- e^+$, respectively.
The present bound on $\lambda_{312}$ comes from measurements on $R_{\tau\mu}=\Gamma(\tau\to \mu \nu \bar{\nu})/\Gamma(\mu\to e\nu \bar{\nu})$~\cite{Ledroit:1998}: $\lambda_{312}<0.062 \cdot \frac{m_{\tilde{\mu}_R}}{100\text{ GeV}}$.
We show two curves for this bound corresponding to sfermion masses of 1 TeV and 2 TeV, respectively.
Similar to the sensitivity plot for $\textbf{B1}$, here we find that the CiADS-BDE can exclude values of the two RPV couplings up to two orders of magnitude beyond the present bounds.

Finally, in the bottom plot, we present the sensitivity reach of the CiADS-BDE to $\textbf{B3}$ in the plane $\lambda_{121}/m^2_{\tilde{f}}$ vs.~$m_{\tilde{\chi}^0_1}$, since in this benchmark scenario we assume there is only one RPV coupling switched on that induces simultaneously the production and decay of the light bino, in the signal processes $\mu^+\to \tilde{\chi}^0_1 \bar{\nu}_e e^+$ and $\tilde{\chi}^0_1\to \overset{(-)}{\nu}_\mu e^+ e^-$.
The existing constraints on $\lambda_{121}$ are obtained from both considerations of charged-current universality (C.C.~universality) and the uncertainty of the measured decay width of $\mu^\pm \to e^\pm+$ invisible ($\sigma_{\Gamma(\mu^\pm\to e^\pm+\text{ invisible})}$).
The C.C.~universality requires that $\lambda_{121}<0.049\cdot \frac{m_{\tilde{e}_R}}{100\text{ GeV}}$~\cite{Barger:1989rk}, and we display this bound for two choices of the sfermion masses, 1 TeV and 2 TeV.
The bound from $\sigma_{\Gamma(\mu^\pm\to e^\pm+\text{ invisible})}$, obtained in Ref.~\cite{Dreiner:2023gir}, is on $\lambda_{121}/m^2_{\tilde{f}}$ and is stronger than that from C.~C.~universality for $m_{\tilde{\chi}^0_1}\lesssim 60$ MeV.
We find that for $m_{\mu}\gtrsim m_{\tilde{\chi}^0_1}\gtrsim 20$ MeV, the CiADS-BDE with $E_p=2$ GeV and $r=1.0$ m can exclude values of $\lambda_{121}/m^2_{\tilde{f}}$ up to about one order of magnitude beyond the present bounds, reaching approximately $2\times 10^{-8}$ GeV$^{-2}$.

We may compare these results with those predicted for SHiNESS~\cite{Wang:2024zky}, and conclude that similar results are expected especially for the $r=1.0$ m setups of the CiADS-BDE.

\section{Conclusions}\label{sec:conclusions}

In a previous work we have proposed a new beam-dump experiment to be constructed behind the existing beam dump at the CiADS facility, called the CiADS-BDE~\cite{Chen:2024fzk}.
The phenomenological study on the sensitivity of the CiADS-BDE to long-lived dark photons shows excellent potential of the proposed experiment to probe new parameter regions that are hard to access by other experiments.
In this work, we have focused on further LLP scenarios where the LLPs can arise from decays at rest of $\pi^+$ mesons and $\mu^+$ leptons at the beam dump.
We have comprehensively chosen HNLs, ALPs with electrophilic couplings or LFV couplings, and the light bino in the RPV-SUSY, for sensitivity studies.
These BSM models are strongly motivated for various reasons including explaining the non-vanishing masses of active neutrinos and the muon $g-2$ anomaly, providing a dark-matter candidate, and solving the hierarchy problem.

Following Ref.~\cite{Chen:2024fzk}, we require that the LLPs should decay to a $e^+ e^-$ pair plus anything as the signal channel, and that both the electron and positron should have an energy larger than 17 MeV and their opening angle should be at least $15^\circ$.
With these requirements, we assume the background level is $100\pm 50$ events per year as an order-of-magnitude estimate~\cite{Chen:2024fzk}.
We have then performed MC simulations in order to determine the acceptance of the CiADS-BDE to these LLPs with different masses and decay lengths, where the acceptance includes both the kinematic cuts on the displaced electron and positron and the decay probability of the LLP inside the detector behind the beam dump.

With a POT of the order of magnitude $\mathcal{O}(10^{23})$ per year, the sensitivity reach of the CiADS-BDE, for 5-year operation time, is found to be excellent, probing large, new parts of the parameter space of the considered BSM scenarios.
In the minimal HNL scenarios where the single, kinematically relevant HNL is mixed with either the electron neutrino or the muon neutrino, we find that the CiADS-BDE with the $r=1.0$ m configuration can exclude values of the mixing parameter squared down to $\mathcal{O}(10^{-9}\text{--}10^{-8})$ ($\mathcal{O}(10^{-7})$) in the electron-mixing (muon-mixing) case, beyond the existing bounds by up to 2 (1) orders of magnitude.
For the electrophilic-ALP case, the CiADS-BDE is found to be able to probe values of $c_{ee}/\Lambda$ down to the order of magnitude $\mathcal{O}(10^{-5})$ GeV$^{-1}$ for $m_a$ roughly between 12 MeV and 130 MeV, touching an unique part of the $(m_a, c_{ee}/\Lambda)$ plane.
In the theoretical scenario of an ALP with the couplings $g_{ee}$ and $g_{\mu e}$, our results show that the CiADS-BDE can cover a large parameter region in the plane $(m_a,g_{ee})$ that lies between the existing bounds obtained from considerations of SN1987A cooling and SN1987A decay.
A clear advantage compared to Mu3e~\cite{Knapen:2024fvh} is observed mainly as a result of the large POT at the CiADS facility.
Finally, for the model of a light bino in the RPV-SUSY, we have studied three benchmark scenarios of LNV RPV coupling pairs.
In all these three benchmark scenarios, we find that the CiADS-BDE can exclude values of the RPV couplings re-scaled by sfermion masses squared orders of magnitude beyond the present bounds.

Compared to another proposed experiment, SHiNESS, to be operated in the vicinity of the target at the ESS, we conclude that the CiADS-BDE shows similar expected performance in testing the BSM scenarios with the corresponding LLPs.

Our work demonstrates the promising prospect of the proposed CiADS-BDE in searching for LLPs predicted in various BSM models, and motivates further studies in this direction for the experiment.

\acknowledgments

YZ is supported by the National Natural Science Foundation of China (Grant No.~12475106) and the Fundamental Research Funds for the Central Universities (Grant No.~JZ2023HGTB0222).
LW Chen is supported by the National Natural Science Foundation of China (Grant No.~12105327) and the Guangdong Basic and Applied Basic Research Foundation (Grant No.~2023B1515120067).

\appendix

\bibliographystyle{JHEP}
\bibliography{refs}

\end{document}